# Toward Trustworthy Short-Range Forecasts with AFNO: From Skill Metrics to Conservation Checks

Akshay Sunil[1], B Deepthi[2*], Muhammed Rashid[3]

[1]Department of Civil Engineering, Indian Institute of Technology Bombay, Powai, Mumbai 400076, India

[2]Kerala State Council for Science, Technology and Environment, Thiruvananthapuram, India

[3]Centre for Climate Studies, Indian Institute of Technology Bombay, Powai, Mumbai 400076, India

*Corresponding author: E-mail: deepthibhadran2@gmail.com

## Abstract

Data driven weather models now approach traditional numerical weather prediction (NWP) skill at short to medium lead times, but their dynamical consistency during autoregressive rollout remains uncertain. This study evaluates a trained Adaptive Fourier Neural Operator (AFNO) for short range regional forecasting using ERA5 reanalysis (0.25° grid, 6 hourly) over 10 to 60°N and 160 to 100°W during 2019 to 2023 (training 2019 to 2022; evaluation 2023). The model predicts five atmospheric variables specific humidity, temperature, zonal wind, meridional wind, and geopotential height at six pressure levels (250 to 850 hPa), producing 6 hourly forecasts to 72 hours via autoregressive rollout. Forecast skill is quantified using anomaly correlation coefficient (ACC), normalized RMSE (NRMSE), and Nash Sutcliffe efficiency (NSE). Beyond statistical skill, we assess physical self-consistency using finite difference diagnostics derived from predicted fields, including the domain maximum horizontal divergence and a normalized horizontal momentum residual calibrated using a 200 member hindcast sample. Across lead times, maximum horizontal divergence remains below $2\times10^{-5}$ s$^{-1}$, and normalized residuals are generally within calibrated bounds, with brief exceedances at mid lead times in upper and mid tropospheric levels. Overall, the results indicate that AFNO can deliver accurate and computationally efficient short-range forecasts while maintaining encouraging consistency with key dynamical diagnostics during multi day autoregressive prediction. The proposed joint skill plus physics evaluation framework supports more trustworthy validation of neural operator forecasts and motivates extensions to longer lead times, additional variables, and hybrid training with explicit physical constraints.

## 1. Introduction

The ability to predict atmospheric variability across different spatial and time scales is essential in today's weather and climate science. Numerical Weather Prediction (NWP) models, which directly incorporate the basic equations of fluid dynamics and thermodynamics, have changed operational forecasting since they appeared in the mid-20$^{th}$ century (Lorenc, 1986; Kalnay, 2003; Stensrud, 2009; Bauer et al., 2015). The ongoing advancement of dynamical cores, data assimilation methods, and improved computing power has gradually enhanced forecast accuracy over recent decades. However, reliable predictions beyond 10 to 15 days remain a challenge. The chaotic nature of the atmosphere (Lorenz, 1963), along with uncertainties in initial conditions, physical models, and boundary factors, limits the reliability of weather forecasts. As a result, the performance of NWP models significantly declines over longer timeframes, making large ensembles and probabilistic approaches necessary (Palmer, 2019; Schultz et al., 2021).

The challenges go beyond just chaos. High-resolution global models that try to resolve convective processes and multiscale interactions demand a lot of computing power. Even with exascale computing, operational centers must choose between resolution, ensemble size, and forecast length. As a result, there has been more focus on complementary methods that take advantage of progress in machine learning (ML) and deep learning (DL) to imitate, speed up, or improve dynamical models (Ham et al., 2019; Reichstein et al., 2019; Janiesch et al., 2021; Ren et al., 2021; Schultz et al., 2021; Tercan and Meisen, 2022; Rolnick et al., 2022). These methods can make good use of the vast observational and reanalysis datasets, providing quick results with less computational effort compared to traditional dynamical simulations.

Early uses of ML in atmospheric science focused on statistical downscaling, bias correction, and parameterization emulation (Vandal et al., 2017; Sachindra et al., 2018; Rasp et al., 2018; Cho et al., 2020; Wang and Tian, 2022; Hosseinpour et al., 2025). Recently, researchers have looked into deep learning models like Convolutional Neural Networks (CNNs), Convolutional LSTMs (ConvLSTMs), and Transformer models for direct spatiotemporal forecasting (Krizhevsky et al., 2012; Shi et al., 2015; Weyn et al., 2020; Li et al., 2021; Jiang et al., 2022; Bi et al., 2023; Sunil et al., 2025). These models have shown promising skill for short- to medium-term forecasts. They also demonstrate deep learning's ability to capture complex nonlinear relationships in atmospheric data. However, many of these models struggle to capture global teleconnections and long-range dependencies, which are essential for longer forecasts (Salman et al., 2018; Guo et al., 2024)

In this context, operator learning has become a promising approach. Instead of learning connections between finite-dimensional state vectors, operator networks aim to approximate connections between function spaces. This makes them especially suitable for systems described by partial differential equations (PDEs) (Lynch, 2008; Müller and Scheichl, 2014; Lu et al., 2021). The Fourier Neural Operator (FNO), introduced by Li et al. in 2021, marked a significant advancement by using spectral representations to learn PDE solutions across different parameter ranges. Building on this groundwork, the Adaptive Fourier Neural Operator

(AFNO) (Guibas et al., 2021; Kurth et al., 2022; Eyring et al., 2024) added adaptive mechanisms for spectral attention. This improvement enhances efficiency and scalability when dealing with large geophysical datasets. These advancements have sparked new interest in using neural operators for weather and climate forecasting.

Initial results have been impressive. AFNO and related architectures have been included in models like FourCastNet, which showed leading performance for global medium-range weather forecasts with much lower computational costs than traditional NWP models (Pathak et al., 2022; Charlton-Perez et al., 2024; Liu et al., 2024). Graph neural networks and Transformer-based methods have also performed well in global forecasting (Scarselli et al., 2009; Zhou et al., 2020; Wu et al., 2021; Keisler, 2022), but AFNO has excelled in combining spectral efficiency and scalability.

Despite these successes, a critical question remains largely unexplored: to what extent do neural operator models preserve the governing physics when forecasts are extended to longer timeframes, specifically three days or more? Unlike numerical weather prediction models, which maintain mass, momentum, and energy through the numerical breakdown of governing equations, data-driven models depend only on empirical training. When forecasts are rolled forward iteratively, even small deviations can build up. This accumulation can lead to systematic biases, violations of conservation laws, or instability. Previous deep learning architectures have shown such problems, where long-term forecasts drift from typical climatological patterns and lose physical consistency (Beucler, et al., 2021, Kashinath et al., 2021). While AFNO's spectral formulation offers some structural benefits, its ability to maintain physical accuracy during long-term iterative forecasts still needs thorough evaluation.

The importance of this question extends beyond academic interest. Reliable longer-range predictions are critical for applications such as agriculture, water resources management, energy planning, and disaster risk reduction. For neural operator models to be used in decision-making at these scales, it is essential to assess whether their forecasts remain consistent with fundamental physical constraints. Models that produce visually realistic fields while exhibiting systematic departures from basic balances, such as hydrostatic equilibrium, potential vorticity conservation, or energy closure, may lead to misleading interpretations. Consequently, the trustworthiness of machine learning based forecasting systems cannot be established through traditional skill metrics alone, such as anomaly correlation coefficient (ACC) or root mean square error (RMSE), but also requires diagnostic evaluation of physical consistency. In this study, physical laws are not explicitly enforced during model training; instead, their consistency is assessed a posteriori through diagnostics computed from the predicted fields during autoregressive rollout.

The present study applies a deep learning-based neural operator framework, specifically the Adaptive Fourier Neural Operator (AFNO), for short-term atmospheric forecasting. The model is developed to predict five core atmospheric variables: specific humidity, temperature, zonal wind, meridional wind, and geopotential height, across six standard pressure levels (250, 300, 350, 400, 500, and 850 hPa). The prediction is performed over a spatial domain covering 10°N to 60°N latitude and 160°W to 100°W longitude, using ERA5 reanalysis data from 2019 to 2023. The model is trained on data from 2019 to 2022 and tested on 2023. The forecasts are generated at 6-hour intervals and evaluated over a 72-hour lead time using autoregressive rollout. The prediction skill is assessed through three statistical metrics: anomaly correlation coefficient (ACC), Nash–Sutcliffe efficiency (NSE), and normalized root mean square error (NRMSE). In addition to statistical evaluation, the study introduces a physics-based validation of the forecasts by computing residuals of the mass continuity and momentum equations using finite difference approximations. Hence, the study fills the gap by evaluating how reliable AFNO forecasts are when extended to long prediction periods (3 days or more). So, mainly this study combines statistical checks with physics-based evaluations to provide a complete view of both predictive accuracy and physical correctness. This approach adds to the existing research on physics-informed machine learning in geosciences, showing both the benefits and drawbacks of using operator learning for long-range forecasting.

The rest of this manuscript is organized as follows. Section 2 describes the study area and the datasets used for model development. Section 3 presents the details of the AFNO architecture, data preprocessing, and experimental design. Section 4 outlines the statistical evaluation and physics-based validation framework, including the diagnostics used to assess mass and momentum conservation. Section 5 presents the results and discussion, focusing on spatial prediction accuracy and physical consistency across different pressure levels. Finally, Section 6 summarizes the main conclusions and gives directions for future research.

**2. Study Area and Data Used**

**2.1 Study Area**

This study looks at a spatial area that stretches from 10°N to 60°N latitude and 160°W to 100°W longitude. This region covers a large part of North America and the nearby eastern Pacific Ocean. It includes the western and central United States, parts of southern Canada, and northern Mexico, along with ocean areas that impact weather patterns across the continent. The chosen region reflects a range of weather conditions across different climate zones, land features, and atmospheric settings (Kalnay et al., 1996; Mesinger et al., 2006).

The southern part of the area includes subtropical regions, such as parts of Mexico and the U.S. Southwest, which have warm, dry climates. As the area goes northward, it crosses mid-latitude zones like the Great Plains and the Pacific Northwest, where strong seasonal changes and large weather systems like fronts and jet streams are common. The inclusion of eastern Pacific waters is especially significant, as many weather systems begin offshore before moving inland (Trenberth and Hurrell, 1994).

This region is important for both weather and its economic impact. It often sees various severe weather events, including winter storms, atmospheric rivers, droughts, and severe storms. Additionally, the area features varied landscapes, from coastal plains and basins to complex mountain ranges like the Rockies. These features create challenges for weather prediction and modelling accuracy (Ham et al., 2019).

Figure 1 displays the study area as a bold red rectangle on a map of the continental United States. The map shows coastlines, national borders, and state lines, providing a clear visual reference. The figure indicates that the area covers much of the western and central U.S. It also extends westward into the Pacific Ocean. This design allows the model to consider atmospheric influences from upstream, which are important for accurate medium-range forecasting.

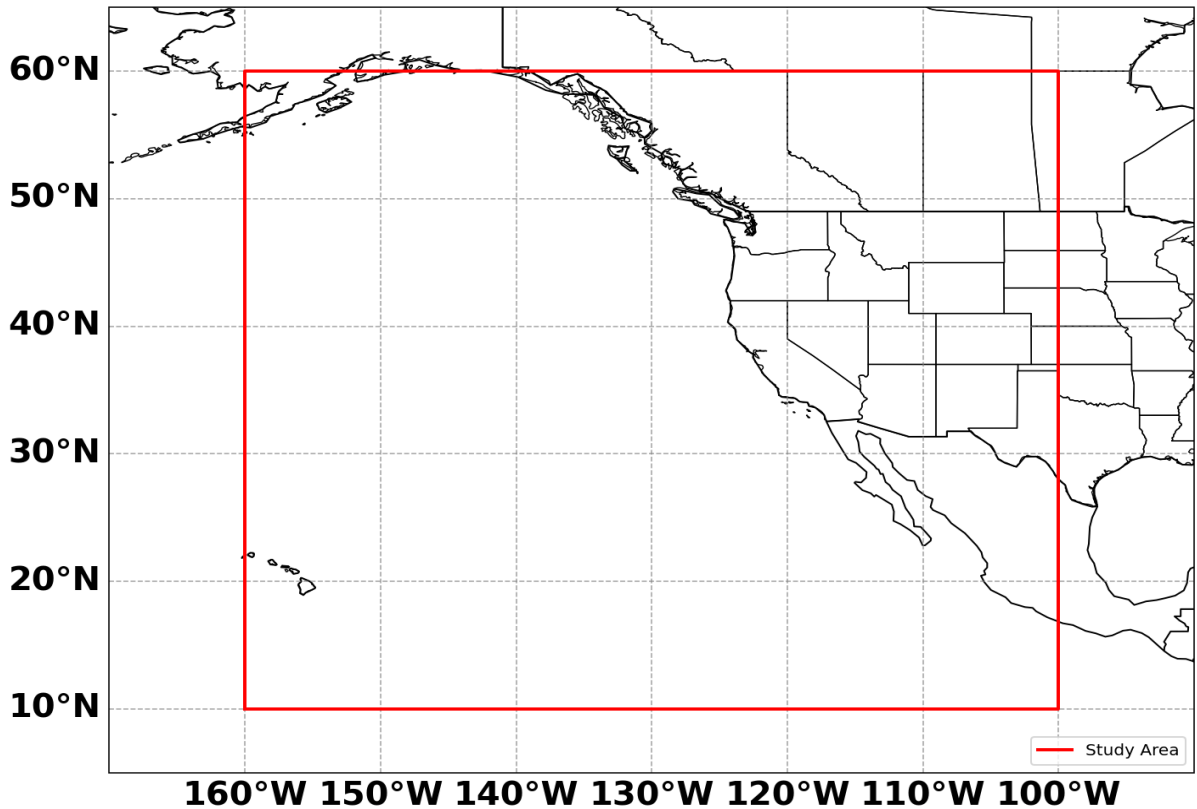

**Figure 1.** Study domain (red box), 10°–60°N and 160°–100°W, spanning the eastern North Pacific and the western–central United States.

**2.2. Data Used**

This study uses atmospheric data from the ERA5 reanalysis dataset provided by the European Centre for Medium-Range Weather Forecasts (ECMWF). ERA5 offers high-quality global weather data at a spatial resolution of 0.25° × 0.25°. This data gives detailed insights into atmospheric patterns and processes. To balance computational efficiency with wide coverage, the analysis focuses on a regional area extending from 10°N to 60°N latitude and from 160°W to 100°W longitude. This region includes North America and nearby oceans, where significant weather events like atmospheric rivers often occur.

The dataset includes five key atmospheric variables measured at six pressure levels: 250, 300, 350, 400, 500, and 850 hPa. This provides a complete vertical profile from the upper troposphere, around 10 km in altitude, to the lower troposphere, approximately 1.5 km in altitude. The variables are geopotential height to identify pressure systems and circulation patterns, temperature for thermal structure analysis, specific humidity for moisture content assessment, and zonal and meridional wind components to characterize atmospheric flow dynamics. Data is available every 6 hours, providing four observations daily. This frequency captures the changing nature of weather systems while keeping computational needs manageable. The dataset covers five years, from 2019 to 2023. The first four years, 2019 to 2022, are used for model training, while the last year, 2023, is reserved for model testing.

## 3. Methodology

This study presents a deep learning approach using AFNO to forecast multivariate atmospheric fields and to evaluate the physical consistency of the generated forecasts based on the fundamental equations governing atmospheric motion. The methodology involves data acquisition and preprocessing, model development, prediction and post-processing, statistical evaluation, and physics-based validation.

### 3.1. Data Acquisition and Preprocessing

The input data are taken from the ERA5 global reanalysis dataset. Five atmospheric variables are selected for forecasting: specific humidity, temperature, zonal wind, meridional wind, and geopotential height. Each variable is sampled at six pressure levels: 250, 300, 350, 400, 500, and 850 hPa. These levels span the lower to upper troposphere, capturing key dynamical layers relevant for synoptic-scale circulation.

All variables are aligned in time to ensure consistency. Each atmospheric field is normalized at every time step using z score normalization, such that the model operates on standardized spatial anomalies rather than absolute values. For each variable, the spatial mean (μ) and spatial standard deviation (σ) are computed over the domain at a given time step, and the field is standardized as

$$x' = \frac{x - \mu}{\sigma} \qquad (1)$$

where x denotes the original field value and x′ is the normalized variable. This normalization removes the instantaneous spatial mean and rescales variability, allowing the model to focus on the evolution of relative spatial patterns and anomalies while reducing sensitivity to seasonal amplitude differences and large scale background states.

### 3.2. Problem Formulation

The forecasting task is defined as a supervised learning problem. For each input at time $t$, the model predicts the atmospheric state at time $t + \Delta t$, where the forecast horizon $\Delta$ is six hours. This setup creates a set of input-output pairs that show the short-term changes in atmospheric conditions. Each sample has thirty channels, which correspond to five variables measured at six pressure levels.

### 3.3. AFNO Model Architecture and Training

The AFNO model maps spatial fields to the frequency domain using a Fourier transform, applies learned spectral filters, and reconstructs forecasts in physical space via an inverse transform, allowing it to capture both local and long-range dependencies efficiently. It is implemented in PyTorch and trained with mean squared error loss,

$$MSE = \frac{1}{N}\sum_{i=1}^{N}(y_i - \hat{y}_i)^2 \tag{2}$$

where $y_i$ and $\hat{y}_i$ are observed and predicted values at $N$ spatial points. Training and validation sets are formed from a split of the available data, model selection is based on minimum validation error, and input–output tensors are stored as memory mapped arrays for efficient data access. As a neural operator that learns mappings between function spaces rather than fixed size vectors, AFNO can represent global atmospheric dependencies while retaining the computational advantages of spectral processing.

### 3.4. From Vector-to-Vector to Function-to-Function Mapping

Traditional neural networks learn mappings between finite-dimensional vectors, such as transforming an input vector x ∈ Rn to output vector y ∈ Rm. In atmospheric forecasting, this approach requires discretizing continuous fields at fixed resolutions, limiting generalization across different grid sizes. AFNO instead learns an operator G: X → X that maps continuous atmospheric state functions to their temporal evolution, where X represents a function space (typically Sobolev space). For atmospheric state $u_t(x)$ at spatial location x and time t, the operator predicts:

$$u_{t+\Delta t} = G_\theta(u_t) \tag{3}$$

where θ represents learnable parameters. This formulation enables resolution-independent predictions and captures physical dynamics that span multiple spatial scales simultaneously.

### 3.5. Spectral representation and computational efficiency

AFNO gains efficiency by operating in the spectral domain. For an atmospheric field $u(x, y)$ on a grid of size $H \times W$, where $H$ and $W$ are the numbers of grid points in the vertical and horizontal directions, direct global interactions in physical space scale as $O((HW)^2)$. In the frequency domain, convolution reduces to elementwise multiplication, lowering the complexity to $O(HW \log(HW))$. The two-dimensional discrete Fourier transform decomposes the field into frequency components:

$$\hat{u}(k_x k_y) = \sum_{i=0}^{H-1}\sum_{j=0}^{W-1} u(i,j) exp\left(-2\pi i \left(\frac{k_x \cdot i}{H} + \frac{k_y \cdot j}{W}\right)\right) \tag{4}$$

where $k_x$ and $k_y$ index discrete wavenumbers and $i$ is the imaginary unit. Low wavenumbers represent large scale features such as planetary waves, while high wavenumbers capture fine scale structure, matching the multiscale nature of atmospheric dynamics.

### 3.6. Layer-by-Layer Information Flow

AFNO processes atmospheric fields through a sequence of transformer-like blocks, each implementing a complete cycle of spatial-to-spectral-to-spatial transformations. For input tensor u(l) at layer l with shape [Batch, Channels, Height, Width], the forward pass executes the following sequence:

**Step 1: Input Normalization**: Layer normalization stabilizes training by normalizing channel statistics across spatial dimensions, preventing covariate shift as gradients flow through deep networks.

$$u^{(l)}_{norm} = \frac{u^{(l)} - \mu}{\sqrt{\sigma^2 + \epsilon}} \qquad (5)$$

**Step 2: Fourier Transform:** Fourier transform converts normalized spatial fields to frequency domain. For real-valued inputs, conjugate symmetry reduces storage to [B, C, H, W/2+1] complex values.

$$\hat{u}^{(l)} = FFT_{2D}\left(u^{(l)}_{norm}\right) \qquad (6)$$

**Step 3: Adaptive Spectral Filtering**: This is the core learnable component. Each frequency mode undergoes non-linear transformation followed by soft thresholding, adaptively emphasizing informative frequencies while suppressing noise:

$$SoftThreshold(z, \lambda) = sign(z).\max(|z| - \lambda, 0) \qquad (7)$$

This soft-thresholding operation introduces structured sparsity in the spectral domain, automatically identifying which frequency components contribute most to prediction accuracy.

**Step 4: Inverse Transform:** Filtered spectral features are converted back to spatial domain for further processing:

$$u^{(l)}_{spatial} = IFFT_{2D}(\hat{u}^{(l)}) \qquad (8)$$

**Step 5: Residual Connection and ML:** A feedforward network with residual connection refines spatial representations and enables gradient flow:

$$u^{(l+1)} = u^{(l)} + MLP\left(u^{(l)}_{spatial}\right) \qquad (9)$$

AFNO is designed to respect the multiscale and time dependent nature of atmospheric dynamics. In the spectral domain it decomposes energy across wavenumbers into three bands corresponding to planetary, synoptic, and mesoscale ranges, using cutoff wavenumbers that define low, intermediate, and high frequency content. Each band is processed by a dedicated spectral filter, and learned attention weights combine the band specific outputs so that the model can emphasize large scale geopotential patterns while still representing fine scale moisture and wind variability. For forecasting, AFNO is applied autoregressively: a seventy-two-hour forecast at six hourly resolution is obtained through twelve sequential applications $u^{(n)} = G_\theta(u^{(n-1)})$. To help control error growth and preserve temporal coherence the network uses learnable temporal embeddings that encode forecast step, hour of day, and day of year, which modulate the spectral processing and allow the model to adapt to diurnal and seasonal variations without explicit physical parameterizations.

The AFNO architecture represents a fundamental reimagining of neural network design for spatiotemporal prediction. By operating in the spectral domain, incorporating multiscale decomposition aligned with atmospheric physics, and maintaining global connectivity through efficient FFT operations, AFNO achieves competitive forecast skill with orders of magnitude fewer computational resources than traditional numerical weather prediction. The architecture's approximately 45 million parameters learn implicit atmospheric dynamics from data while maintaining physical consistency through spectral constraints, temporal embeddings, and adaptive filtering mechanisms. This combination of computational efficiency, physical interpretability, and strong empirical performance positions AFNO as a promising foundation for next-generation weather forecasting systems.

### 3.7. Statistical Evaluation

After training, the model is used to generate predictions on the test data. The outputs are initially obtained in normalised form. These predictions are then transformed back into physical units using the stored time-dependent means and standard deviations. The predicted fields and the corresponding ERA5 observations are saved for further evaluation.

The accuracy of the forecasts is evaluated using four statistical metrics. The mean absolute error (MAE) is calculated as:

$$MAE = \frac{1}{N}\sum_{i=1}^{N}|y_i - \hat{y}_i| \tag{10}$$

The normalized root mean square error (NRMSE) is given by:

$$NRMSE = \frac{\sqrt{\frac{1}{N}\sum_{i=1}^{N}(y_i - \hat{y}_i)^2}}{\max(y) - \min(y)} \tag{11}$$

The anomaly correlation coefficient (ACC), which quantifies the similarity between observed and predicted anomalies, is expressed as:

$$ACC = \frac{\sum_{i=1}^{N}(y_i - \bar{y})(\hat{y}_i - \bar{\hat{y}})}{\sqrt{\sum_{i=1}^{N}(y_i - \bar{y})^2}\sqrt{\sum_{i=1}^{N}(\hat{y}_i - \bar{\hat{y}})^2}} \tag{12}$$

The Nash-Sutcliffe efficiency (NSE) is defined as:

$$NSE = 1 - \frac{\sum_{i=1}^{N}(y_i - \hat{y}_i)^2}{\sum_{i=1}^{N}(y_i - \bar{y})^2} \tag{13}$$

Here, $y_i$ and $\hat{y}_i$ denote the observed and predicted values respectively, $\bar{y}$ and $\bar{\hat{y}}$ represent their respective means, and $N$ is the total number of values.

### 3.8. Physics-Based Validation

In addition to the statistical evaluation, a physics-based validation is conducted to assess whether the AFNO-generated forecasts adhere to the governing atmospheric dynamics. The trained model is applied in an autoregressive manner from 0 to 72 hours, with forecasts generated at 6-hours intervals. At each step, the predicted fields are used as inputs for the subsequent forecast, thereby simulating continuous forecasting. For each lead time and across six isobaric pressure levels ranging from 250 hPa to 850 hPa, the horizontal wind components (u and v), temperature (T), specific humidity (q), and geopotential height are extracted.

Diagnostics of mass continuity and horizontal momentum balance are then computed using centered finite difference approximations. The momentum conservation equation is expressed as:

$$\frac{\partial v}{\partial t} + (v.\nabla)v + fk \times v = -\nabla\Phi - \frac{1}{\rho}\nabla p + \nu\nabla^2 v \tag{14}$$

Where v = (u,v) is the horizontal wind vector, $f$ is the Coriolis parameter, $\rho$ is air density, $p$ is pressure and $\nu$ is the kinematic viscosity.

The mass continuity equation is given by:

$$\frac{\partial \rho}{\partial t} + \nabla.(\rho v) = 0 \tag{15}$$

All spatial derivatives and divergence terms are computed using central finite difference approximations. The residuals of these equations are evaluated at each grid point and over all forecast lead times.

In practice, due to the use of pressure level data and the absence of vertical velocity and density tendency terms, mass consistency is evaluated using the horizontal divergence of the predicted wind field as a necessary but not sufficient condition for mass conservation. This physical validation step provides additional insights into the dynamical consistency of the predictions. A small residual indicates that the model-generated forecasts are consistent with the governing physical laws. A gradual increase in residuals over time reveals the lead time beyond which the predictions begin to diverge from physically realistic behaviour. Figure 2 illustrates the overall workflow followed in this study. The top panel (a) shows the data preprocessing and model training pipeline. Atmospheric variables including specific humidity, temperature, zonal wind, meridional wind, and geopotential height are extracted from the ERA5 reanalysis dataset at six pressure levels ranging from 250 hPa to 850 hPa. These variables are normalized at each time step and used to construct input-output image pairs for training and testing. The Adaptive Fourier Neural Operator (AFNO) model is then trained using these data. The bottom panel (b) outlines the prediction and validation stages. The trained AFNO model generates forecasts, which are then de-normalized for evaluation. The predicted fields are compared against observed fields using statistical metrics such as anomaly correlation coefficient (ACC), normalized root mean square error (NRMSE), and Nash–Sutcliffe efficiency (NSE). Additionally, physics-based diagnostics including mass continuity and horizontal momentum conservation are computed to evaluate the physical consistency of the forecasts.

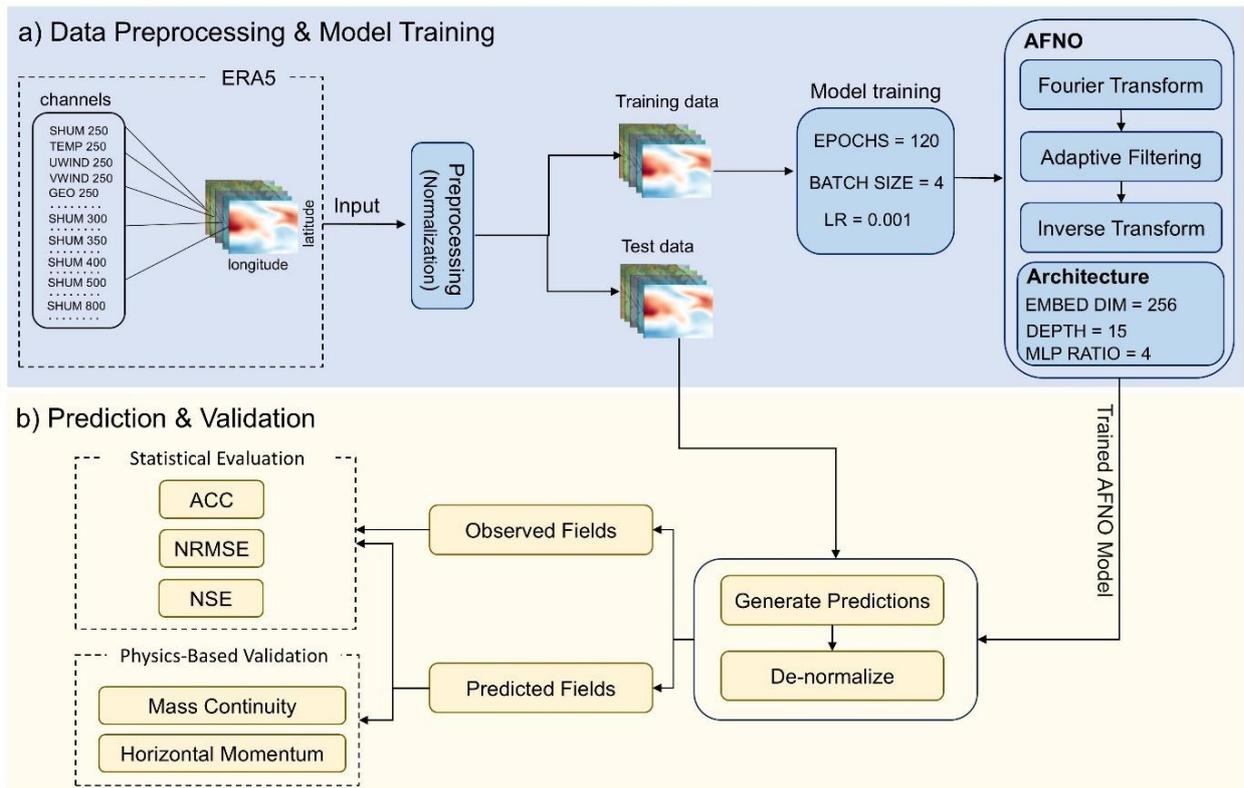

**Figure 2:** Overview of the methodology adopted in this study, including data preprocessing, AFNO model training, prediction, and both statistical and physics-based evaluation.

## 4. Results and Discussion

To evaluate the spatial accuracy of the model, the predicted atmospheric fields at a specific forecast time are compared with reference fields from the ERA5 reanalysis. Figure 3 shows a visual comparison between the predicted and observed data for each of the five variables: specific humidity, temperature, zonal wind, meridional wind, and geopotential height. These variables are assessed at six pressure levels: 250, 300, 350, 400, 500, and 850 hPa. For clarity, the variables are represented in the figure using the following labels: SHUM (specific humidity), TEMP (temperature), UWIND (zonal wind), VWIND (meridional wind), and GEO (geopotential height). The model's predictions closely match the observed fields across all variables and levels. For specific humidity, the model captures the main moisture distribution patterns and the spatial gradients linked to large-scale transport features, especially in the upper atmosphere. For temperature, the predicted fields accurately represent the spatial distribution of thermal structures, including areas of warm and cold anomalies. The horizontal temperature gradients remain consistent, and no significant deviation is seen in the upper and mid-tropospheric layers.

In the wind fields, both the zonal and meridional components show strong agreement with the reanalysis data. The model effectively captures the orientation, intensity, and position of jet streams and wind shear zones, particularly at higher altitudes like 250 and 300 hPa. It reproduces the spatial patterns of upper-level winds well, and the overall flow structure matches the observed circulation features. For geopotential height, the predicted fields mirror the large-scale pressure distributions with a high level of spatial coherence. The model maintains the ridges and troughs in the geopotential height fields, including the gradient structures that are important for understanding the dynamic behaviour of the atmosphere. While some lower-level predictions show slight smoothing, the overall structural integrity of the fields remains strong.

Overall, the predicted fields exhibit strong spatial correlation with the observed values for all five variables and at all pressure levels. A visual inspection confirms that the model can create realistic atmospheric fields that align with reanalysis data. This shows that the model effectively captures the multivariate and multilevel structure of the atmosphere at individual time steps.

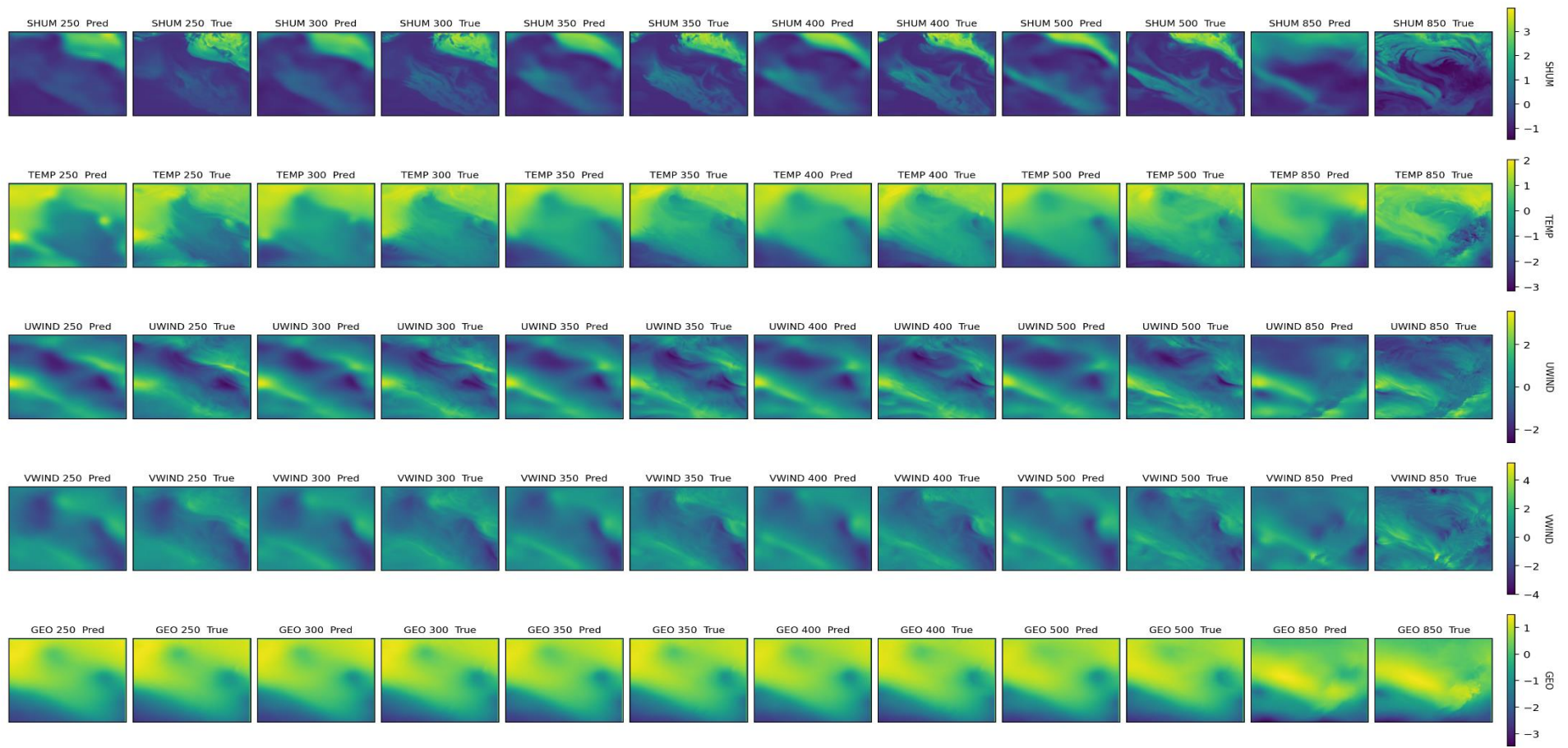

**Figure 3.** Comparison between AFNO model predictions and ERA5 reanalysis fields at a representative forecast time. Each row corresponds to one atmospheric variable, and each pair of columns shows the predicted and observed fields at different pressure levels ranging from 250 hPa to 850 hPa.

## 4.1. Spatial Evaluation Metrics

The accuracy of the spatial forecasts from the AFNO model is evaluated using three standard statistical measures: anomaly correlation coefficient, normalized root mean square error, and Nash-Sutcliffe efficiency. Each measure is calculated at every grid point and shown for all five atmospheric variables and six pressure levels. These spatial maps help assess the model's predictive performance and its ability to represent consistent atmospheric structures throughout the vertical column.

### 4.1.1 Anomaly Correlation Coefficient (ACC)

To quantify agreement between predicted and observed spatial anomaly patterns, the anomaly correlation coefficient is computed at each grid point. Figure 4 shows spatial maps of the anomaly correlation coefficient for all variables and pressure levels. By correlating departures from the mean rather than absolute values, this metric emphasizes pattern and phase alignment of the forecast anomalies relative to the reference field and is therefore well suited for evaluating structural fidelity in gridded atmospheric forecasts.

Temperature and geopotential height exhibit consistently high anomaly correlation coefficient values, typically exceeding 0.9 over most of the domain, indicating strong skill in representing large scale thermal structure and synoptic scale circulation features. The wind components also show high correlations, particularly at upper levels where the flow is more coherent and dominated by well-organized jet and wave patterns. A modest reduction is evident at 850 hPa, consistent with increased near surface variability, stronger influence of boundary layer processes, and enhanced sensitivity to local gradients.

Specific humidity shows comparatively lower anomaly correlation coefficient values, reflecting the greater intermittency and small-scale variability of moisture fields. Nevertheless, the model retains useful skill in capturing the dominant moisture transport patterns, with stronger agreement at mid and upper levels than near the lower troposphere. Overall, the spatial anomaly correlation coefficient results indicate that AFNO maintains robust phase alignment of predicted anomalies across variables and pressure levels, with expected reductions for moisture and near surface flow where predictability is intrinsically lower.

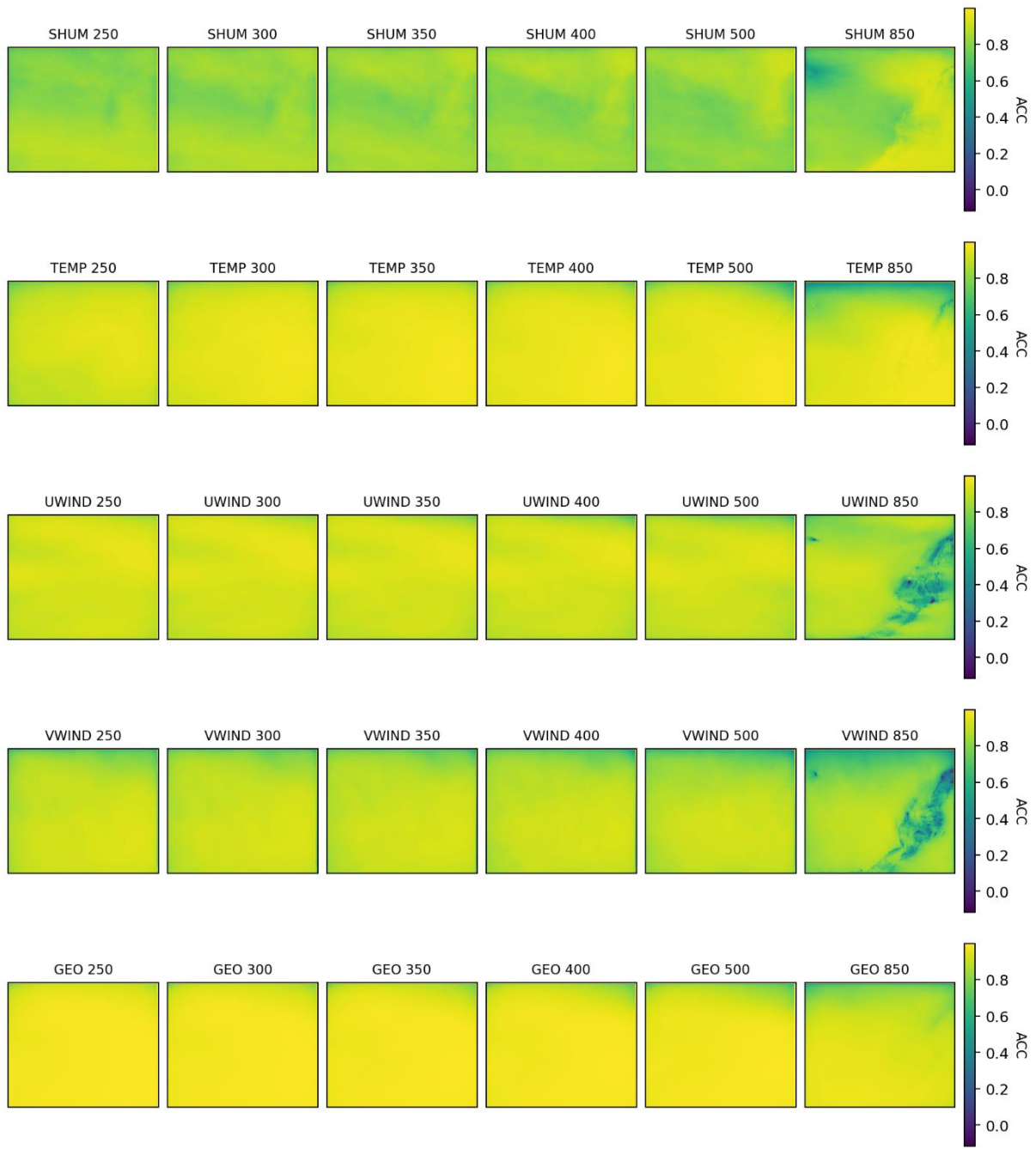

**Figure 4:** Spatial distribution of the anomaly correlation coefficient (ACC) between predicted and observed atmospheric fields at different pressure levels. Each row corresponds to a different variable, while columns represent the six pressure levels used in the model.

Table 1 shows the average anomaly correlation coefficient for geopotential height (GEO), temperature (TEMP), zonal wind (UWIND), meridional wind (VWIND), and specific humidity (SHUM) at 250, 300, 350, 400, 500, and 850 hPa. ACC values reflect how well forecasts match reference anomalies. Higher values mean better alignment. Geopotential height and temperature display the strongest correlation skill overall. GEO reaches 0.977 at 250 hPa and stays high through the mid-troposphere, at 0.959 at 500 hPa, before dropping to 0.916 at

850 hPa. Temperature hits its peak at 350 hPa with a value of 0.954 and stays above 0.94 between 300 and 500 hPa, but shows a lower correlation at 850 hPa with a value of 0.895. Wind skill is generally lower than GEO and TEMP and decreases significantly at the lowest level. At 850 hPa, ACC falls to 0.819 for UWIND and 0.780 for VWIND, compared to about 0.92 to 0.93 in the upper levels. Specific humidity has the lowest correlation performance, with ACC values between 0.823 and 0.834 across the levels tested. This shows limited phase agreement for moisture variability.

**Table 1:** Spatially averaged anomaly correlation coefficient for geopotential height (GEO), temperature (TEMP), zonal wind (UWIND), meridional wind (VWIND), and specific humidity (SHUM) at 250, 300, 350, 400, 500, and 850 hPa.

| Pressure(hPa) | GEO | TEMP | UWIND | VWIND | SHUM |
|---|---|---|---|---|---|
| 250 | 0.977 | 0.920 | 0.928 | 0.892 | 0.831 |
| 300 | 0.974 | 0.947 | 0.928 | 0.893 | 0.832 |
| 350 | 0.971 | 0.954 | 0.924 | 0.890 | 0.832 |
| 400 | 0.968 | 0.953 | 0.920 | 0.884 | 0.828 |
| 500 | 0.959 | 0.943 | 0.909 | 0.867 | 0.823 |
| 850 | 0.916 | 0.895 | 0.819 | 0.780 | 0.834 |

### 4.1.2. Nash–Sutcliffe Efficiency (NSE)

The NSE measures how well the AFNO model predicts compared to the climatological mean. Figure 5 shows the distribution of NSE values for five atmospheric variables at six pressure levels. Values close to one mean the predicted values are very close to the observed values. In contrast, values near zero indicate that the model does not perform better than the average of the observations.

For all variables and levels, the AFNO model achieves very high NSE scores. Most grid points have values above 0.998. This indicates the model's strong ability to capture both the magnitude and spatial differences of atmospheric fields. Geopotential height and temperature fields show particularly uniform and nearly perfect NSE values. This emphasizes the model's skill in forecasting smoothly changing thermodynamic features.

The wind components also show high NSE values at all levels. However, there is a slight decrease in efficiency at 850 hPa in areas with strong spatial changes. These minor differences may result from increased variability driven by the surface and non-linear effects in the lower atmosphere. Specific humidity has the lowest NSE values, but they are still strong, matching observed data well throughout the column.

In summary, the spatial maps of NSE confirm that the AFNO model delivers accurate, dependable forecasts that maintain strong performance across dynamic and thermodynamic variables. The consistent high NSE values across the domain highlight the model's potential for weather forecasting and climate evaluation.

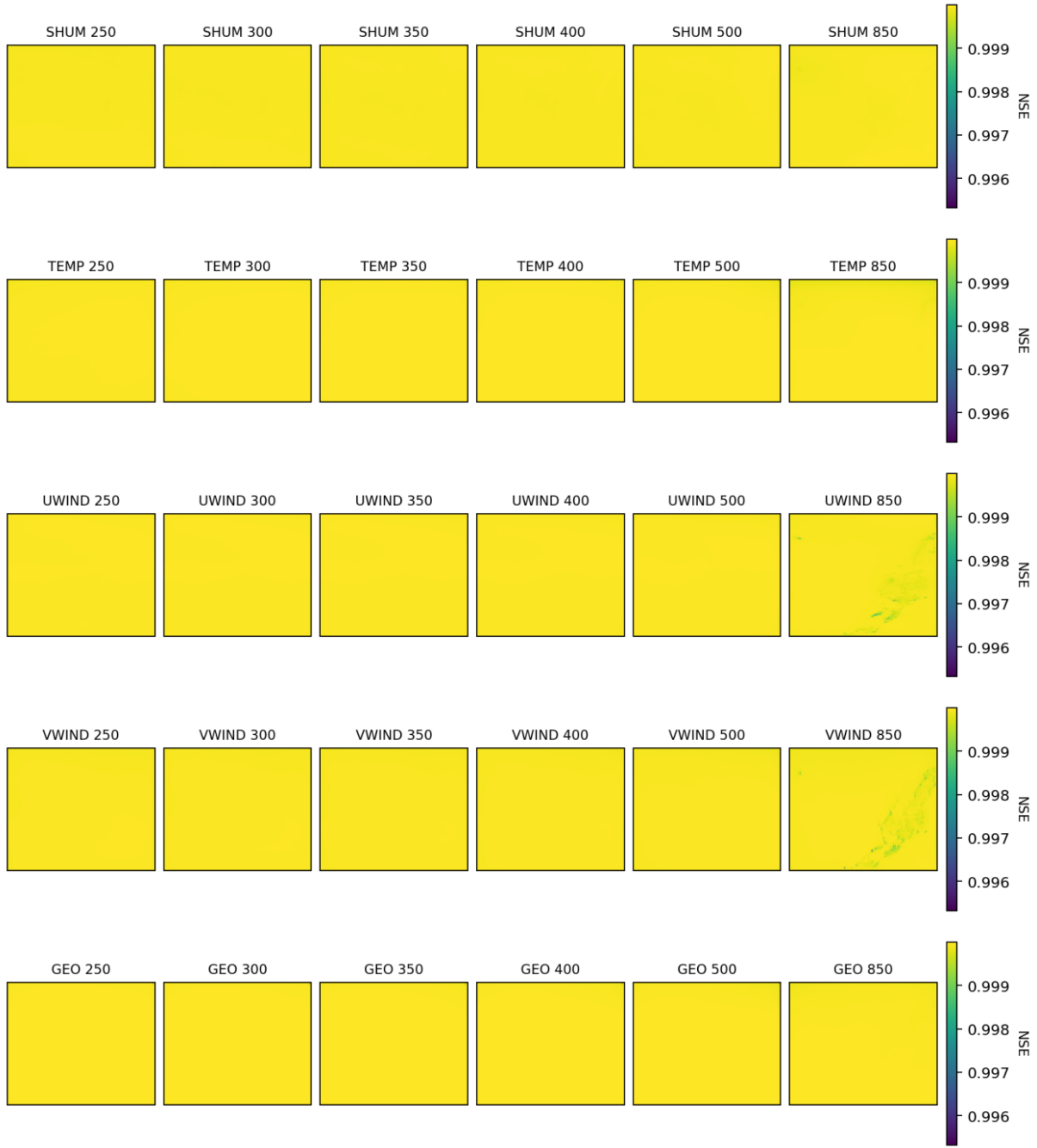

**Figure 5.** Spatial distribution of the Nash–Sutcliffe Efficiency (NSE) computed between predicted and observed atmospheric fields. Each panel corresponds to a specific variable and pressure level.

Table 2 shows the spatially averaged Nash-Sutcliffe efficiency for the five variables at six different pressure levels. NSE values are nearly 1 for all variables and levels, ranging from about 0.99994 to 0.99999. This confirms very high overall skill across the domain. Geo and Temp consistently have NSE values above 0.99997, with peaks nearing 0.99999 in the upper and mid-troposphere. Winds experience a slight drop at 850 hPa, where NSE falls to 0.999946 for UWIND and 0.999936 for VWIND. Specific humidity stays around 0.999955 to 0.999959 across levels.

**Table 2:** Spatially averaged NSE for geopotential height (GEO), temperature (TEMP), zonal wind (UWIND), meridional wind (VWIND), and specific humidity (SHUM) at 250, 300, 350, 400, 500, and 850 hPa.

| Pressure(hPa) | GEO | TEMP | UWIND | VWIND | SHUM |
|---|---|---|---|---|---|
| 250 | 0.999994 | 0.999978 | 0.999980 | 0.999970 | 0.999957 |
| 300 | 0.999993 | 0.999986 | 0.999980 | 0.999970 | 0.999957 |
| 350 | 0.999992 | 0.999988 | 0.999979 | 0.999969 | 0.999957 |
| 400 | 0.999991 | 0.999987 | 0.999978 | 0.999968 | 0.999956 |
| 500 | 0.999988 | 0.999985 | 0.999976 | 0.999965 | 0.999955 |
| 850 | 0.999976 | 0.999971 | 0.999946 | 0.999936 | 0.999959 |

**4.1.3. Normalised Root Mean Square Error (NRMSE)**

Figure 6 shows the spatial distribution of NRMSE for all five variables across six pressure levels. Most variables show very low NRMSE values, typically below 0.02 across the area. Specific humidity has slightly higher NRMSE values, especially at 250 hPa. This is consistent with the challenge of capturing moisture variability in the upper troposphere. Still, the spatial pattern of NRMSE stays consistent, with few localized differences.

Temperature, geopotential height, and both wind components display uniformly low error values at all levels. This reaffirms the model's ability to correctly capture both dynamic and thermodynamic structures. The performance is particularly stable at mid and lower levels, such as 500 and 850 hPa. As expected, there is a slight rise in NRMSE near the surface at the 850 hPa level for wind and specific humidity, but the values are still within acceptable limits. This indicates strong generalization capability.

The low NRMSE values overall, when viewed alongside the ACC and NSE results, suggest that the AFNO model not only captures the phase alignment and timing variations of the fields but also maintains high accuracy in magnitude.

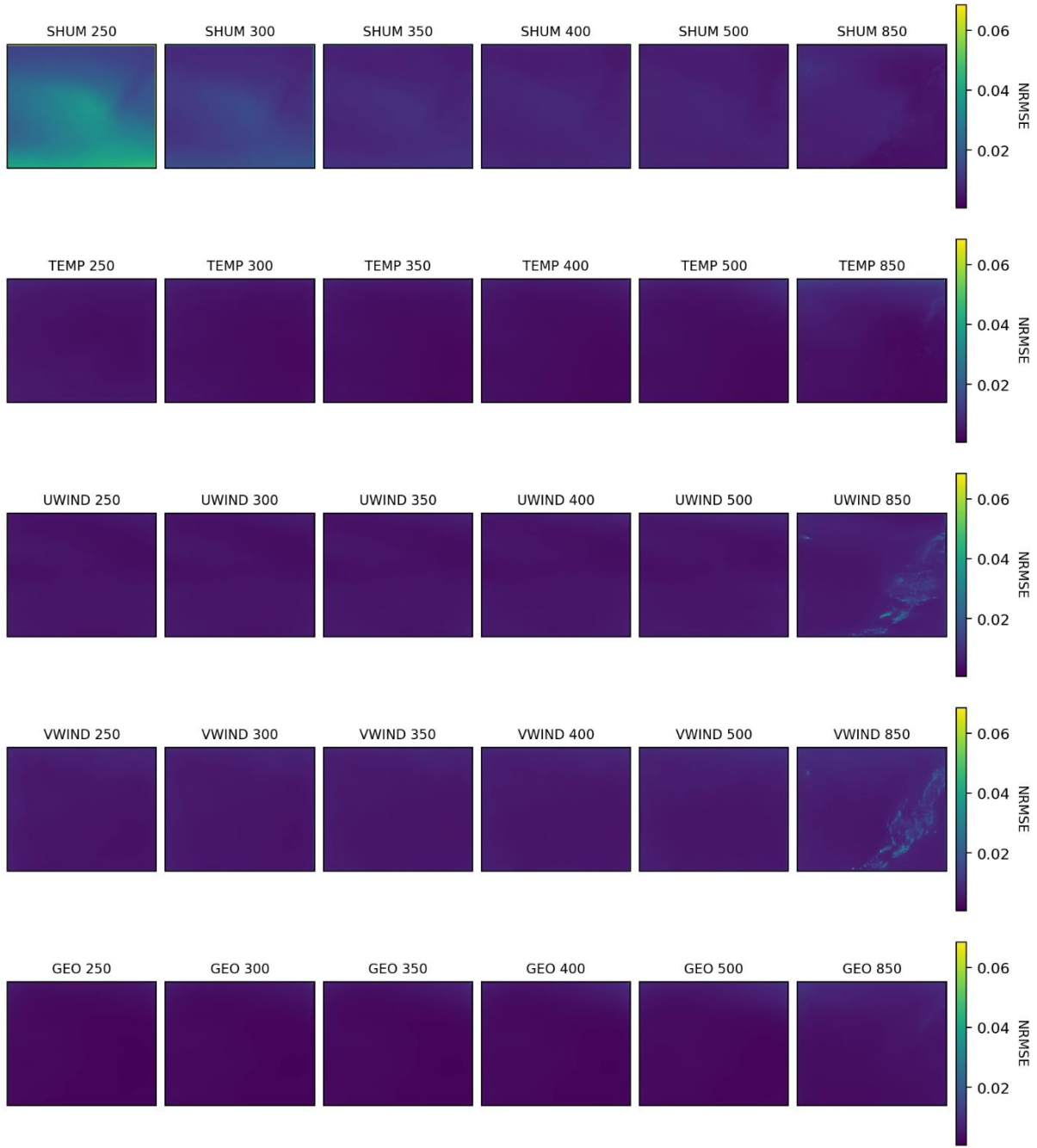

**Figure 6.** Spatial distribution of normalised root mean square error (NRMSE) between predicted and observed fields for each variable at six pressure levels

Table 3 summarizes the spatially averaged normalized root-mean-square error for the five variables at six different pressure levels. Geopotential height exhibits the lowest errors overall, increasing gradually from 0.00238 (250 hPa) to 0.00308 (500 hPa) and reaching 0.00466 at 850 hPa. Temperature errors are similarly low, with minimum NRMSE at 350 hPa (0.00340) and higher values toward both upper and lower levels (up to 0.00482 at 850 hPa). Wind errors increase toward the lower troposphere and peak at 850 hPa (0.00692 for UWIND and 0.00757 for VWIND), indicating the largest low-level amplitude errors among the dynamical variables. Specific humidity shows the strongest vertical contrast, with a pronounced upper-level error maximum at 250 hPa (0.02730), while mid-to-lower levels remain substantially smaller

(approximately 0.00629–0.01283). Overall, Table 3 highlights that the largest normalized errors are associated with upper-tropospheric humidity and near-surface winds, whereas GEO and TEMP maintain low normalized errors throughout the column.

**Table 3:** Spatially averaged NRMSE for geopotential height (GEO), temperature (TEMP), zonal wind (UWIND), meridional wind (VWIND), and specific humidity (SHUM) at 250, 300, 350, 400, 500, and 850 hPa.

| Pressure(hPa) | GEO | TEMP | UWIND | VWIND | SHUM |
|---|---|---|---|---|---|
| 250 | 0.00238 | 0.00455 | 0.00443 | 0.00539 | 0.02730 |
| 300 | 0.00250 | 0.00365 | 0.00442 | 0.00538 | 0.01283 |
| 350 | 0.00262 | 0.00340 | 0.00450 | 0.00544 | 0.00871 |
| 400 | 0.00275 | 0.00342 | 0.00461 | 0.00553 | 0.00747 |
| 500 | 0.00308 | 0.00365 | 0.00487 | 0.00582 | 0.00688 |
| 850 | 0.00466 | 0.00482 | 0.00692 | 0.00757 | 0.00629 |

### 4.2. Physics-Based Validation of Model Forecasts

This section evaluates the dynamical consistency of the AFNO forecasts using physics-based diagnostics computed directly from the predicted fields. Standard verification metrics quantify forecast skill but do not explicitly test whether the model maintains physically plausible balance relationships during iterative autoregressive rollout. To address this, a consistency assessment is carried out using the mass continuity and horizontal momentum equations. The goal is not to claim strict conservation, but to quantify whether departures from these balances remain bounded and non-growing with lead time under repeated model application.

To assess model behaviour under dynamically active conditions, we perform a multi event diagnostic analysis using three atmospheric river cases on January 4, 2023, March 14, 2023, and December 4, 2023. These events were selected to span distinct synoptic configurations and moisture transport intensities, providing complementary test scenarios for evaluating the consistency of AFNO forecasts across different regimes.

Figure 7 presents the atmospheric river detection results for each case using an objective diagnostic based on integrated vapor transport and integrated water vapor. For each event, three panels are shown: integrated vapor transport magnitude with wind vectors, a binary atmospheric river mask derived from combined thresholds (integrated vapor transport (IVT) greater than 250 kg m$^{-1}$ s$^{-1}$ and integrated water vapor(IWV) greater than 20 mm), and integrated water vapor with the 20 mm contour overlaid. This representation highlights both the strength of moisture transport and the spatial coherence of the associated moisture plume.

The January case exhibits a well-defined southwest to northeast oriented atmospheric river over the eastern Pacific with peak integrated vapor transport exceeding 800 kg m$^{-1}$ s$^{-1}$. The March case shows a broader and more zonally oriented moisture corridor, while the December case features a comparatively narrow and intense filament with a concentrated integrated vapor transport core. Together, these detections confirm that the selected dates correspond to atmospheric river conditions and provide the spatial context for the initial states used in the subsequent autoregressive forecast rollouts and physical consistency diagnostics.

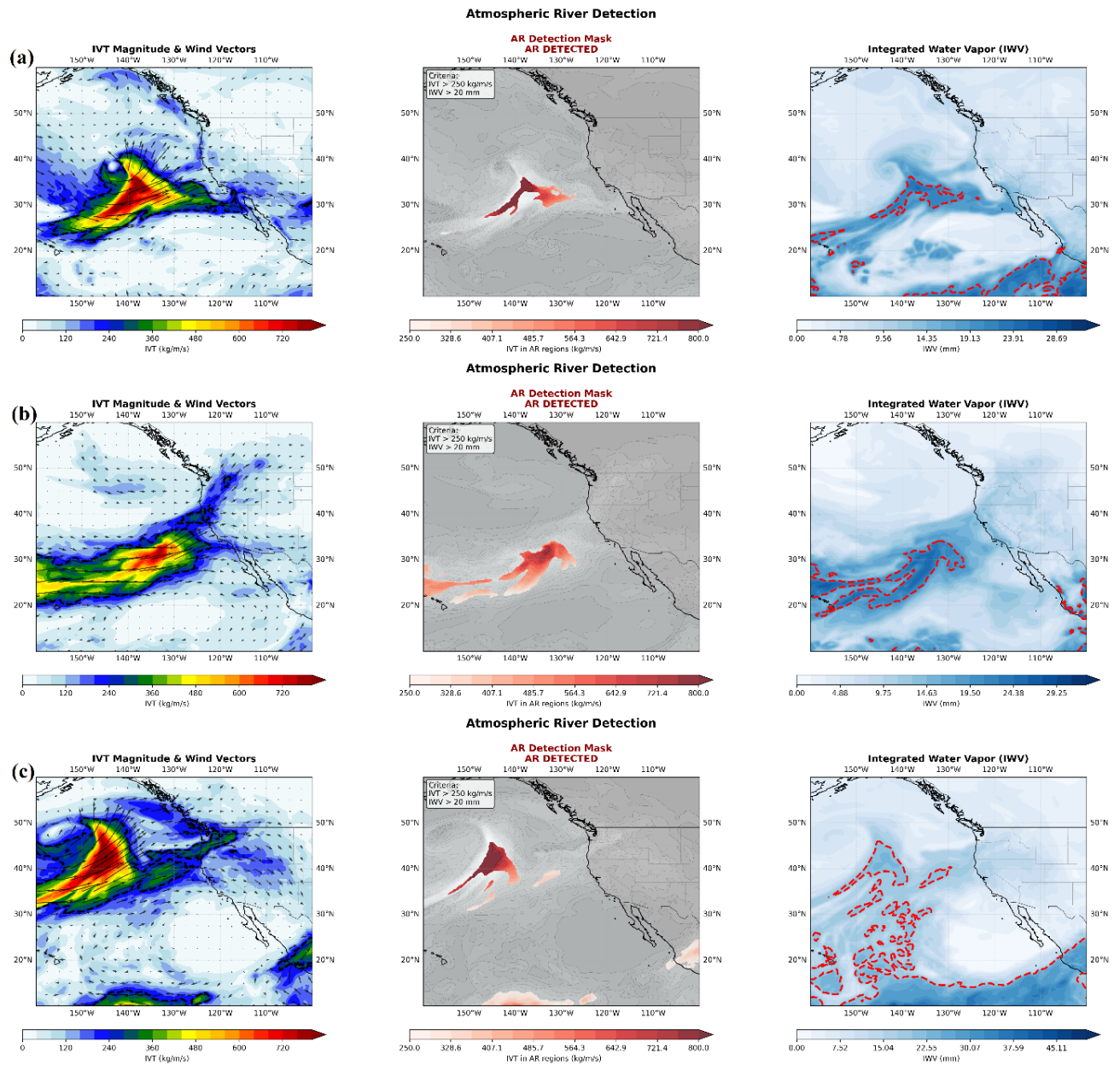

**Figure 7:** Atmospheric River detection plots for (a) January 4, 2023, (b) March 14, 2023, and (c) December 4, 2023. Each row shows (left) IVT magnitude (kg/m·s$^{-1}$) with wind vectors, (center) detected AR regions based on IVT > 250 kg/m·s$^{-1}$ and IWV > 20 mm, and (right) IWV (mm) fields with the 20 mm threshold contour in red.

Building on the identification of the three atmospheric river events, a detailed physical validation is performed to assess how well the AFNO model preserves fundamental conservation principles under these dynamically active conditions. The selected cases—January 4, March 14, and December 4, 2023—serve as representative test scenarios for autoregressive rollout of the model over a 72-hour forecast period. Using the detected AR structures as initial conditions, the analysis tracks the evolution of key dynamical diagnostics, including mass divergence and momentum residuals, across the forecast horizon and multiple pressure levels.

To evaluate the performance of the trained AFNO model across different weather events, a multi-event diagnostic analysis is conducted. Three independent atmospheric events are selected: January 4, 2023, March 14, 2023, and December 4, 2023. For each event, the model is initialized using the observed state and run continuously for 72 hours with a 6-hour interval, simulating a forecasting scenario.

Figure 8 shows the evolution of mass and momentum consistency diagnostics at three pressure levels (250, 500, and 850 hPa) as a function of forecast lead time. For each event, the plotted threshold values represent the mean computed over all available forecast initializations and verification times across the full dataset period (training plus testing years). The top row shows the mean of the domain maximum absolute horizontal divergence of the predicted wind field, used here as a diagnostic proxy for mass consistency. The bottom row shows the mean normalized momentum residual, computed from finite difference estimates of the resolved horizontal momentum balance and scaled using the calibrated reference range.

Across lead times and levels, the mean mass divergence remains below the calibrated threshold (red dashed line), indicating stable mass consistency in the diagnostic sense adopted here. In contrast, the mean normalized momentum residual exhibits stronger level dependence and approaches or exceeds the threshold at intermediate lead times, particularly at upper and mid tropospheric levels. Importantly, the exceedances are not associated with monotonic growth with lead time, suggesting that momentum balance is the more sensitive diagnostic under autoregressive rollout while overall dynamical behaviour remains stable at short range lead times.

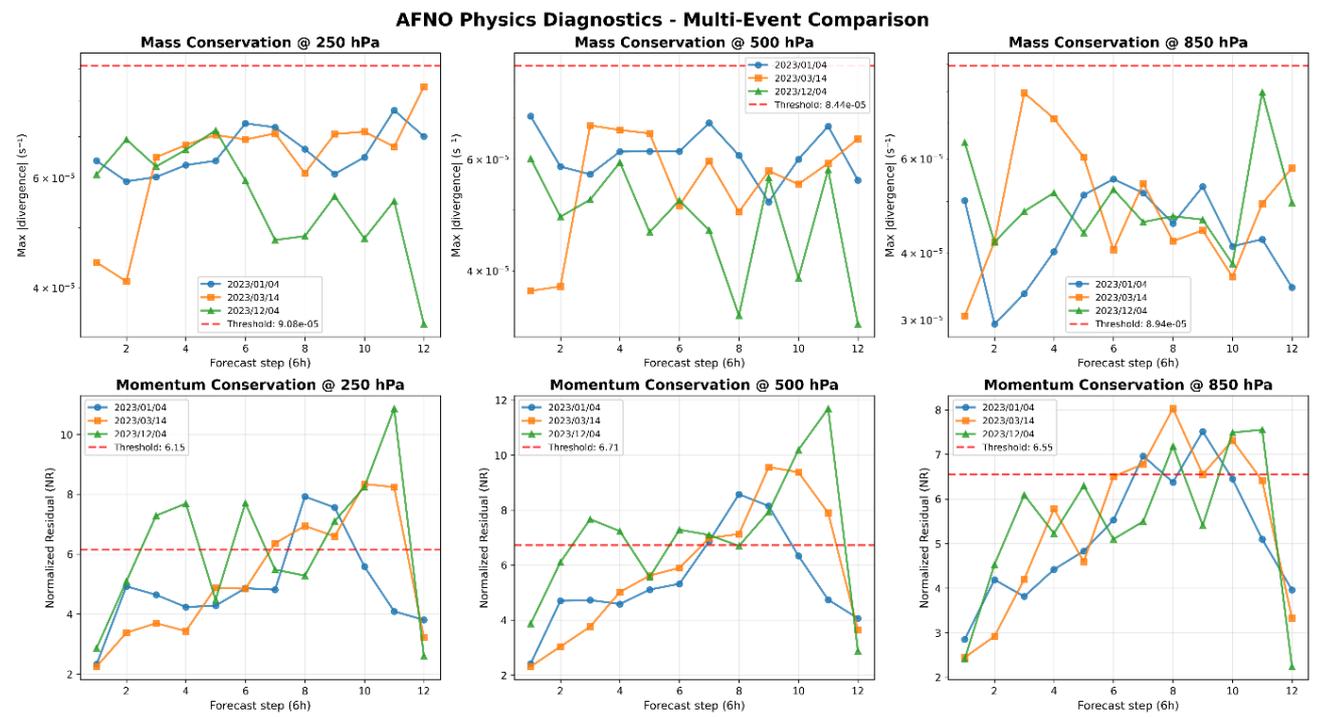

**Figure 8:** Multi-event evaluation of physical consistency in AFNO forecasts. The top row shows the maximum absolute divergence of horizontal wind vectors ($|\nabla \cdot v|$) over 72-hour forecasts for three atmospheric events at 250, 500, and 850 hPa, indicating mass conservation performance. The bottom row displays normalized residuals of the horizontal momentum equations, quantifying adherence to dynamical balance. Each color represents a different atmospheric event—January 4 (blue), March 14 (orange), and December 4 (green). Red dashed lines denote calibrated thresholds for each level derived from climatological analysis.

## 5. Conclusions

This study applied the Adaptive Fourier Neural Operator model for short range regional atmospheric forecasting using multivariate inputs from the ERA5 reanalysis dataset. The model predicts five key atmospheric variables, specific humidity, temperature, zonal wind, meridional wind, and geopotential height, across six standard pressure levels (250, 300, 350, 400, 500, and 850 hPa). This vertical coverage spans the upper to lower troposphere and supports representation of both dynamical and thermodynamic structures relevant to synoptic scale evolution.

Forecast skill was evaluated using the anomaly correlation coefficient, Nash Sutcliffe efficiency, and normalized root mean square error. Across variables, pressure levels, and the spatial domain, the forecasts show strong agreement with ERA5 in both spatial structure and magnitude, indicating that the model captures key spatiotemporal dependencies over short lead times under autoregressive rollout.

Beyond statistical performance, the study introduced a physics based diagnostic evaluation of forecast consistency using finite difference measures derived from predicted fields. Mass consistency was assessed using the domain maximum horizontal divergence of the predicted wind field, which remained below calibrated thresholds across pressure levels and forecast steps in the evaluated cases. In contrast, the normalized momentum residual showed intermittent and event dependent threshold exceedances at intermediate lead times, particularly in the upper and mid troposphere, but did not display systematic growth with lead time or evidence of instability. These diagnostics are computed a posteriori from model outputs and do not reflect explicit enforcement of governing equations during training or inference.

Overall, the combined statistical and physics-based evaluation framework provides a more comprehensive basis for judging the trustworthiness of data driven forecasts than skill metrics alone. The results indicate that the model can deliver accurate and computationally efficient short-range forecasts while maintaining encouraging dynamical consistency during autoregressive prediction, with momentum balance emerging as the more sensitive constraint under active regimes. Future work should extend this framework to longer lead times, additional variables and vertical levels, broader event sets, and hybrid training strategies that incorporate explicit physical constraints or balance aware loss terms.